\documentclass{iaus}
\usepackage{graphicx}

\title[Masses of black holes] 
{Uncertainties of the masses of black holes \\ and Eddington ratios in AGN}

\author[S.~Collin]   
{Suzy Collin}

\affiliation{LUTH, Observatoire de Paris-Meudon, 5 Place Janssen, 92140 Meudon, France\\[\affilskip] 
email: suzy.collin@obspm.fr}

\pubyear{2006}
\volume{238}  
\pagerange{001--999}
\date{??? and in revised form ???}
\jname{Black Holes: from Stars to Galaxies -- across the Range of Masses}
\editors{V. Karas \& G. Matt, eds.}

\begin{document}

\maketitle

\begin{abstract}
Black hole masses in Active Galactic Nuclei have been determined in 35
objects through reverberation mapping of the emission line region. I
mention some uncertainties of the method, such as the ``scale factor''
relating the Virial Product to the mass, which depends on the unknown
structure and dynamics of the Broad Line Region. 

When the black hole masses are  estimated indirectly using the empirical
size-luminosity relation deduced from this method, the uncertainties can
be larger, especially when the relation is extrapolated to high and low
masses and/or luminosities. In particular they lead to Eddington ratios
of the order of unity in samples of Narrow Line Seyfert 1. As the
optical-UV luminosity is provided by the accretion disk, the accretion
rates can be determined and are found to be much larger than the
Eddington rates. 

So, accretion must be performed at a super-critical rate through a slim
disk, resulting in rapid growth of the black holes. The alternative is
that the mass determination is wrong at this limit.
\end{abstract}

\firstsection 
\section{Introduction}

It is a paradox that the determination of Black Hole (BH) masses in
Active Galactic Nuclei (AGN) is more difficult than in quiescent
galaxies. Nevertheless, it is absolutely necessary to know the BH masses
at high redshifts, and to understand how BHs grow. BH masses in AGN are
determined through the `virial technique', so-called because it assumes
that the line emission region is gravitationally bound to the BH. Since
it has become an industry and was used in several dozens of papers to
determine the masses of thousands of BHs, I thought that it was
necessary to recall some basic uncertainties plaguing this method. Also
the question of the accretion rates is very important in the context of
BH growth, and it is almost always confused with the question of the
luminosity. 

\section{Virial masses}

In 35 Seyfert  and low redshift quasars, the BH masses have been
determined directly from ``reverberation mapping'' (in the following
these AGN are called ``Reverberation Mapped'', or RM objects). As a
by-product of this method, an empirical relation was found between the
size of the Broad Line Region (BLR) and the optical luminosity, $L_{\rm
opt}$. In all other AGN except one (NGC 4258), the BH masses are
determined indirectly using this empirical relation.

The direct method consists in measuring the time delay $\tau$  between
the continuum and the line variations which respond to them; it gives a
characteristic size of the BLR.  Assuming that the BLR is
gravitationally bound (which is certainly true for the Balmer line
emitting region, cf. Peterson \& Wandel 2000), the mass of the BH,
$M_{\rm{}BH}$,  is then equal to $M_{\rm BH} = f {c \tau \ \delta V^2/
G}$, where  $\delta V$  is the dispersion velocity, and   $f$  a scale
factor. $c \tau \delta V^2/ G$ is called the ``Virial Factor''. The
usual way is to identify $\delta V$  with the FWHM, and to assume
$f=3/4$,  which correspond to an isotropic BLR with a random
distribution of orbits. 

There are strongly debated questions about the method. To quote only a
few:
\medskip

\begin{description}
\item{1.}  What is the best choice for measuring the dispersion
velocity? 
Is it the FWHM (which is used in all works) or  $\sigma_{\rm
line}$, i.e. the second moment of the line profile? There is a large
scatter of the  FWHM/$\sigma_{\rm line}$ ratio, and Peterson {\etal}
(2004) showed that  $\sigma_{\rm line}$ seems more reliable for the
intrinsic dispersion of measurements, but it is generally not measured.

\item{2.} The scale factor $f$ varies among the objects by a factor
three: it is larger than two for RM objects with narrow peaked lines and 
smaller than unity in RM objects with broad flat topped lines. Collin
{\etal} (2006) showed that it depends probably on the Eddington rate and
on the inclination of the BLR (which is most  likely not spherical but
axially symmetric), and that some narrow line objects must be seen at
low inclination and consequently have their masses underestimated by up
to one order of magnitude.

\item{3.}  Is it better to use the RMS or the mean spectrum? 
Peterson {\etal} (2004) reanalysed the data of all RM objects, used
$\sigma_{\rm line}$ instead of the FWHM, RMS spectrum instead of mean
spectrum, and changed the factor $f$, as scaled on the bulge masses by
Onken {\etal} (2004). As a result, they obtained masses {\it
systematically larger} by typically a factor two than those of Kaspi
{\etal} (2000), some of them by factors up to one order of magnitude.
\end{description}
\medskip

Everybody in this audience is aware that there is an empirical relation
discovered with the  RM objects between the size of the BLR and the
optical luminosity (Peterson \& Wandel 1999,  Kaspi {\etal} 2000).  This
relation was revised by Kaspi {\etal} in 2005 and is given now with a
good precision. It is not well understood theoretically, but it allows
to determine $M_{\rm{}BH}$ for single epoch observations, by simply
measuring $L$(opt) (=$\nu L_{\nu}$ at 5100\AA) and the FWHM.  This is
very useful since the reverberation method requires at least several
months of monitoring of a given object to lead to a mass determination.
Once the mass is known,  it is also possible to determine the Eddington
ratio $R_{\rm Edd} = L_{\rm bol}/L_{\rm Edd}$, assuming a bolometric
correction of the order of 10. It has been used for large samples of
quasars, like the SDSS. 

Some studies have shown that the most luminous quasars correspond to
very massive black holes, up to $10^{10}M_{\odot}$ radiating at their
Eddington luminosity. At the other extreme, it led to very small masses,
of the order of a few 10$^{5}M_{\odot}$, in a sample of AGN with small
host galaxies (Barth {\etal} 2005). It is not clear whether the
extrapolation of the relation to large or small luminosities is valid
(cf. Wang \& Zhang 2003 for the low mass side). Moreover one must not
forget that the indirect method eliminates the intrinsic dispersion of
the size-luminosity relation, and that the scale factor is still not
well known and certainly varies among objects. 

\section{The Eddington ratio versus the accretion rate}

Using the RM objects and assuming that the optical luminosity was due to
the accretion disk, Collin {\etal} (2002) showed that a fraction of them
were accreting at super-Eddington rate. Actually the rates were
overestimated, because the Hubble constant was assumed to be 50 and not
70 km s$^{-1}$ Mpc$^{-3}$, and the masses taken from Kaspi {\etal}
(2000) were probably underestimated by a factor two to three. In total
the Eddington ratios were overestimated by a factor of about six, and
the accretion rates in Eddington units had to be also reduced by larger
factors (see below), so almost none of the RM objetcs are now found to
accrete at super-Eddington rates. Nevertheless, applying the same
technique to samples of Narrow Line Seyfert 1 nuclei (NLS1s) where the
BH masses were determined by the indirect Virial method, Collin \&
Kawaguchi (2004) showed that a large fraction of them are accreting at
super-Eddington rate. 
	
\begin{figure}
\begin{center}
 \includegraphics[width=\textwidth]{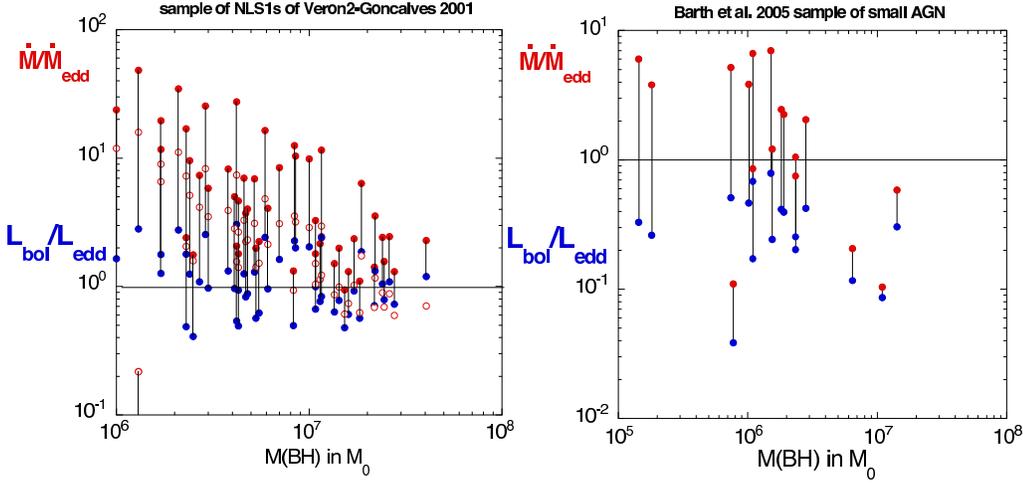}
  \caption{Accretion rates and luminosities in Eddington units for two samples (see the text).}  
  \end{center}
\end{figure}
	
It is easy to understand why the accretion rate in Eddington units can
be much larger than the Eddington ratio. In the optical range, the disk
radiates locally like a black body (this is not true in the UV range,
cf. Hubeny {\etal} 2001). Thus $L_{\rm opt}$ can be expressed as 
$L_{\rm opt}=A\ {\rm cos}(\theta) (M \dot{M})^{2/3}$, where $\theta$ is
the inclination of the normal of the disk with the line of sight, and
$\dot{M}$ is the accretion rate. $A$ is a constant, actually of the
order of unity. It leads to the efficiency $\eta$ for mass-energy
conversion:	
\begin{equation}
\eta \sim 0.006 {{\rm cos}(\theta)^{3/2} C_{\rm bol10} M_6^{3/2} \over \sqrt{R_{\rm Edd}}}
\label{eta}
\end{equation}
where $C_{\rm bol10} \approx 1$ is the observed bolometric correction
divided by 10, and $M_6$ the mass in 10$^6M_{\odot}$. Thus $\eta$ could
be very small for large luminosities and small BH masses. It means that
{\it the accretion rate expressed in Eddington units with a standard
efficiency of 0.1} ($\dot{M}_{\rm Edd}=L_{\rm Edd}/0.1 c^2$) 
{\it is much larger than} $R_{\rm Edd}$. 
This is illustrated in Fig.~1. The figure
on the left (from Collin \& Kawaguchi 2004)  shows the application of
this law to the NLS1 sample of Veron-Cetti {\etal} (2001), and that on
the right to the sample of Barth {\etal} (2005) consisting of small mass
BHs. The red points correspond to $\dot{M}/\dot{M}_{\rm Edd}$, and the
blue ones to $R_{\rm Edd}$. 

One can see that $\dot{M}/\dot{M}_{\rm Edd}$ is typically one order of
magnitude larger than $R_{\rm Edd}$ for the NLS1 sample, while it is
slightly smaller for the Barth sample whose luminosities are smaller. On
the left figure are also shown the values of $\dot{M}/\dot{M}_{\rm Edd}$
computed assuming that a large fraction of the optical luminosity is
provided by an external non-gravitational heating of the disk (cf.
Collin \& Kawaguchi 2004 for a detailed explanation). Though the values
of $\dot{M}/\dot{M}_{\rm Edd}$ are smaller than previously, they remain
nevertheless larger than $R_{\rm Edd}$. 

\begin{figure}
\begin{center}
\includegraphics[width=0.55\textwidth]{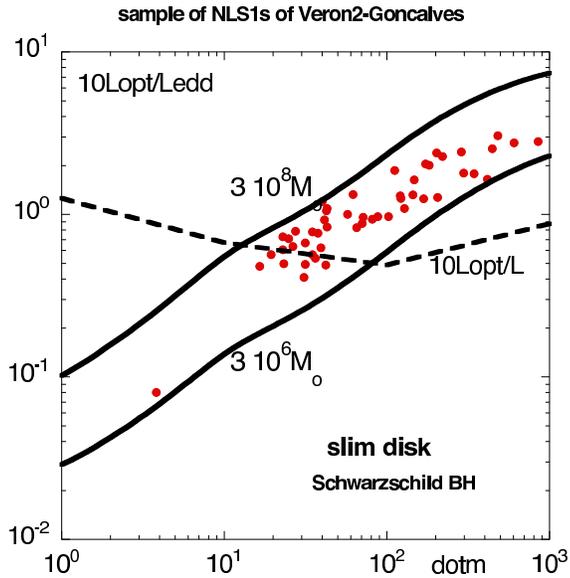}
\hfill
\parbox[b]{0.4\textwidth}{\caption{Observed values for the Veron {\etal} sample 
compared to slim disk parameters.\newline}}
 \end{center}
\end{figure}

Is this conclusion unescapable? Other explanations can be proposed:
\medskip

\begin{description}
\item[1.]{~Super-Eddington accretion rate can be relatively large at a
distance of 100 to 1000 Schwarzschild radii corresponding to the
emission of the optical band, and relativistic outflows can be created
close to the BH. Note however that such outflows, if they exist as
proposed by Gierlinski \& Done (2004), are not super-Eddington.}
\item[2.]{~The optical-UV emission is not due to the accretion disk,
even when taking into account a non-gravitational external heating: but
then to what else (see the discussion in Collin {\etal} 2002)?}
\item[3.]{~The empirical size-luminosity relation may not be valid at
large Eddington ratios and small masses, or the scale factor is much
larger for these objects.} \item[4.]{~Alternatively, we observe really
super-Eddington accretion rates, due to ``slim'' super-Eddington
advective disks (Abramowicz {\etal} 1988).}
\end{description}
\medskip

Indeed  the SED of such disks is in agreement with the relation between
$\dot{M}/\dot{M}_{\rm Edd}$ and  $R_{\rm Edd}$, and with the constancy
of the ratio $L_{\rm bol}/L$opt, as shown on Fig. 2 (cf.  the
description of the slim disk model in Collin \& Kawaguchi 2004).  If
these super-Eddington accretion rates are real and concern not only
nearby quasars, they would have important cosmological consequences.
During their low mass phase, the growth time of the BHs would not be
Eddington but mass supply limited, and can be much smaller than the
Eddington time. Super-Eddington accretion can thus account for the rapid
early growth of BHs. It implies also that the BH/bulge mass relationship
for NLS1s would be more dispersed than in other objects.

\begin{acknowledgments}
I am grateful to T.~Kawaguchi, B.~Peterson, and M.~Vestergaard for many
enlightening discussions on the subject. 
\end{acknowledgments}


\bigskip

\discuss{Debora Dultzin-Hacyan}{Comment: It has been shown by Zhang \& Win that the
$R$--$L$ relation is not valid for dwarf (low $L$) AGN.}

\discuss{Suzy Collin}{You are perfectly right. If the size is larger that
that given by $R$--$L$ relation for small masses, it means that we 
underestimate them. On the other hand, as we have shown (Collin {\etal}
2006) that the scale factor is probably underestimated for the NLS1
class, this could be another reason of underestimating the mass.}


\begin{thebibliography}{}

\bibitem[]{a} Abramowicz M.A., 
Czerny B., Lasota J.-P., \& Szuszkiewicz E. 1988, ApJ, 332, 646 

\bibitem[]{b} Barth A.J., Greene 
J.~E., Ho L.~C. 2005, ApJL, 619, L151

\bibitem[]{c} Collin, S., Boisson, C., 
Mouchet M., Dumont A.-M., Coup{\'e} S., Porquet D., Rokaki E. 
2002, A\&A, 388, 771  

\bibitem[]{d} Collin S.,
Kawaguchi T. 2004, A\&A, 426, 797 

\bibitem[]{e}   Collin S., Kawaguchi T., Peterson B.~M. \& Vestergaard 2006, A\&A, 456, 75

\bibitem[Gierli{\'n}ski \& Done(2004)]{} Gierli{\'n}ski M., Done C. 2004, MNRAS, 349, L7 

\bibitem[]{f} Hubeny, I., Blaes, O., Krolik J.~H., \& Agol E. 2001, ApJ, 559, 680 

\bibitem[]{g}    Kaspi S., Smith P.~S., Netzer H., Maoz D., Jannuzi B.~T., Giveon U. 2000, ApJ, 533, 631 

\bibitem[]{h}    Kaspi S., Maoz D., Netzer H., Peterson B.~M., Vestergaard M., Jannuzi B.~T. 2005, ApJ, 629, 61 

\bibitem[]{i}     Onken C.~A., Ferrarese L., Merritt D., Peterson B.~M., Pogge R.~W., Vestergaard M., Wandel A.\ 2004, ApJ, 615, 645

\bibitem[]{j}    Peterson B.~M., Wandel A.\ 1999, ApJL, 521, L95

\bibitem[]{k}    Peterson B.~M., Wandel A. 2000, ApJL, 540, L13

\bibitem[]{l}     Peterson B.~M. {\etal} 2004, ApJ, 613, 682 

\bibitem[]{m} V{\'e}ron-Cetty, M.-P., V{\'e}ron, P., \& Gon{\c c}alves, A.C. 2001, A\&A, 372, 730 

\bibitem[]{n} Wang, T.-G., \& Zhang, X.-G. 2003, MNRAS, 340, 793 

\end{thebibliography}
\end{document}